\documentclass[fleqn,twoside]{article}
\usepackage{espcrc2}
\usepackage{graphicx}
\usepackage[figuresright]{rotating}

\newcommand{\x}{\mbox{$\vec{x}$}}
\newcommand{\simg}{\rlap{\raise -4pt \hbox{$\sim$}}
                   \raise 3pt \hbox{$>$}}
\newcommand{\siml}{\rlap{\raise -4pt \hbox{$\sim$}}
                   \raise 3pt \hbox{$<$}}

\title{Application of DWF to heavy-light mesons}

\author{N. Yamada\address[RBRC]{RIKEN BNL Research Center, 
        Brookhaven National Laboratory, Upton, NY 11973, USA}%
        ~~for the RBC Collaboration%
        \thanks{We thank RIKEN, Brookhaven National Laboratory and the
                U.S. Department of Energy for providing the facilities
                essential for the completion of this work.
               }
       }
       
\begin{document}

\begin{abstract}
 We consider application of domain wall fermions to quarks with
 relatively heavy masses, aiming at precision calculations of charmed
 meson properties.
 Preliminary results for a few basic quantities are presented.
\end{abstract}

\maketitle

\section{INTRODUCTION}

 After the remarkable accomplishment of the $B$ factories,
 theoretical uncertainty remaining in the decay amplitudes of
 heavy-light mesons is an important obstacle to the stringent test of the
 Standard Model.
 Lattice QCD offers a promising method to reduce this uncertainty in a
 systematic way and numerous attempts have been made to achieve this
 purpose so far \cite{Ryan:2001ej,Yamada:2002wh}.
 While the accuracy we can attain in simulating $b$ quark is still
 limited, mainly due to difficulty in taking a continuum limit, present
 computational resources should make it possible to perform precision
 calculations of charmed hadron properties.
 It is expected that the charm factories will allow us to calibrate
 lattice calculations of the heavy-light system with high precision, and
 the experiences gained in $c$-physics will be advantageous to the
 $b$-physics simulations.

 Domain wall fermions (DWF)
 \cite{Kaplan:1992bt,Shamir:1993zy,Furman:ky} are expected to have only
 a negligible $O(a)$ error as well as better control over chiral
 extrapolations.
 Non-perturbative renormalization, which is essential to high precision
 calculations, also seems to work remarkably well \cite{Blum:2001sr}.
 Simulating light quarks using DWF benefits from these advantages and
 has been successfully used in the light hadron system
 \cite{Blum:2000kn,AliKhan:2000iv,Aoki:2002vt}.
 It is, then, natural to consider its application to the massive quarks.
 As a first step toward the precision calculations, we have started the
 simulation of $D$ mesons using DWF for both heavy and light quarks.

\section{SIMULATION PARAMETERS}

 The first exploratory study of massive DWF \cite{Norman:2003_lat03}
 observed that with an improved gauge action at weaker gauge coupling
 DWF do not show clear failure for $am_q\siml$0.4.
 Based on this observation, we take the DBW2 gauge action
 \cite{Takaishi:1996xj} with $\beta$=1.22 at which the charm quark is
 realized around $am_q\sim$0.4.
 The simulation is carried out using quenched lattices with
 $24^3\times48$.
 We set the extension of the fifth dimension and the domain wall height
 to $L_s$=10 and $aM_5$=1.65, respectively. 
 The light and heavy quark masses range in
 0.008$\le am_{lq}\le$0.040 and 0.1$\le am_{hq}\le$0.5, respectively.
 To avoid the finite size effect in the time direction, all the meson
 two-point functions are calculated under both the periodic and
 anti-periodic boundary conditions and then averaged.
 The gauge field is fixed to the Coulomb gauge in this work, and
 $a^{-1}$=2.86(9) GeV is obtained using $m_\rho$=770 MeV.
 Data presented below is obtained with 20 configurations.

 The magnitude of the explicit chiral symmetry breaking can be probed by
 the residual mass, $m_{\rm res}$, defined by
 \begin{eqnarray}
    a m_{\rm res}
  = \frac{\langle \sum_x J_{5q}^a(\x,t)\pi^a(0)\rangle}
         {\langle\sum_x J_{5 }^a(\x,t)\pi^a(0)\rangle},
 \end{eqnarray}
 where the explicit forms of the operators are found in
 Ref. \cite{Blum:2000kn,Aoki:2002vt},
 and we obtain $a m_{\rm res}=9.72\times 10^{-5}$, which roughly
 corresponds to $m_{\rm res}\sim$ 0.3 MeV.

\section{PRELIMINARY RESULTS}

\begin{figure}[t]
 \hspace*{-4ex}
 \includegraphics[width=5.5cm,angle=-90]{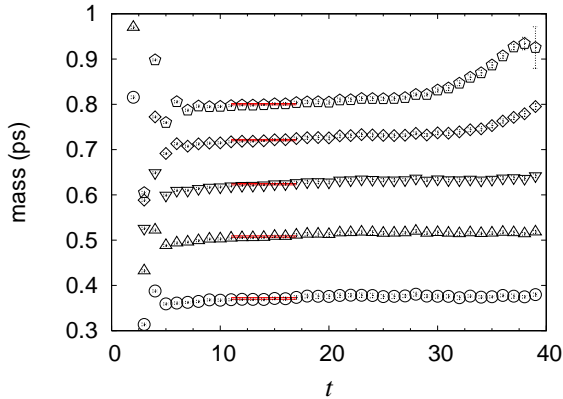}
 \vspace{-3ex}
 \caption{Effective mass plots of the heavy-light pseudo-scalar mesons.
          The fit ranges and the results are indicated by horizontal
          lines.}
 \label{fig:efplt-040-ps}
\end{figure}
 To see if the large mass parameters cause any unexpected behavior, we
 first examine the effective mass plots.
 The plots shown in Figure \ref{fig:efplt-040-ps} are obtained from the
 $\langle A_4 P^\dag\rangle$ correlation functions with wall source and
 point sink, where $A_4$ and $P$ are heavy-light bilinears.
 The light quark mass is fixed to $am_{lq}$=0.040 while the heavy quark
 mass varies from 0.1 (bottom) to 0.5 (top) by 0.1.
 The data with $am_{hq}$=0.4 and 0.5 show a gradual increase at
 $t\simg$30 and it looks getting more conspicuous as $am_{hq}$ gets
 larger, otherwise no odd behavior is seen for all heavy quark masses.
 We repeated the same calculation for one configuration with more
 stringent stopping condition and found that in the heaviest case
 ($am_{hq}$=0.5) the correlation function with the smaller stopping
 conditions differs by only +0.05\% at $t$=20 while the difference
 reaches to +10\% around $t\sim$30.
 Therefore it seems that the increase found in
 Figure \ref{fig:efplt-040-ps} is caused by the loose stopping condition.
 Since all the correlation functions for $t\le$20 show no odd behavior
 and are seen to have a reasonable plateau, we take this range to
 perform a conservative analysis.
 The errors given below are statistical only, unless otherwise
 specified.

\begin{figure}[t]
 \hspace*{-4ex}
 \includegraphics[width=5.5cm,angle=-90]{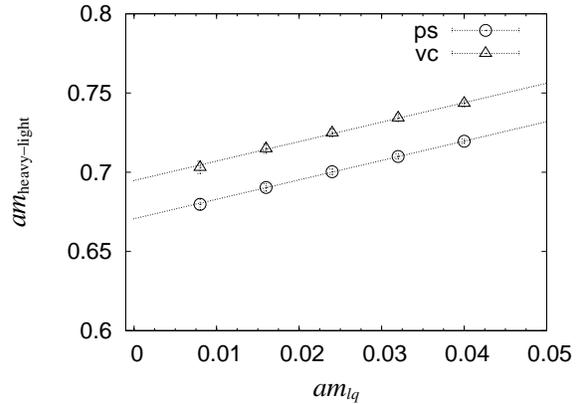}
 \vspace{-3ex}
 \caption{Light quark mass dependences of the heavy-light pseudo-scalar
          (circles) and vector (triangles) meson masses.}
 \label{fig:chiextrp-400-ps}
\end{figure}
 Figure \ref{fig:chiextrp-400-ps} shows the chiral extrapolation of the
 heavy-light pseudo-scalar (circles) and vector (triangles) meson masses
 with $am_{hq}$=0.4.
 Both the linear and quadratic fits describe the data very well, and the
 difference due to the different fits does not exceed 0.5 \% in the
 chiral limit  for the both mesons.
 We choose the linear fit to obtain the heavy-light masses at
 $am_{ud}$=$-am_{\rm res}$ and $am_s$=0.032(2) for later use, where
 $am_s$, corresponding to the strange quark mass, is determined using
 $m_K/m_\rho$ as a representative among other possible inputs.
 To find $am_{\rm charm}$, the pseudo-scalar masses are interpolated to
 the physical value of $m_D$ using a quadratic function of heavy quark
 mass, and we obtain $am_{\rm charm}$=0.380(4).
 The mass splitting of $m_{D_s}-m_D$ is, then, obtained as
 \begin{eqnarray}
  m_{D_{s}} - m_D = 114(6)\ \mbox{ MeV}.
 \end{eqnarray}
 While this value is about 15\% larger than the experimental value
 (99MeV), quenching may account for much of the discrepancy.

 The lattice calculation of the 1S hyperfine splitting has been a long
 standing problem because the lattice results are significantly
 smaller than the experimental values, independent of the heavy quark action
 adopted.
 In this work we obtain
 \begin{eqnarray}
   m_{D^*}   - m_D     &=& 69(6)\ \mbox{ MeV},\\
   m_{D^*_s} - m_{D_s} &=& 70(4)\ \mbox{ MeV}.
 \end{eqnarray}
 Comparing to the experimental results of $m_{D^*}-m_D$=142 MeV and
 $m_{D^*_s} - m_{D_s}$=144 MeV, we see that this problem persists when
 using DWF.

 \begin{figure}[t]
  \hspace*{-4ex}
  \includegraphics[width=5.5cm,angle=-90]{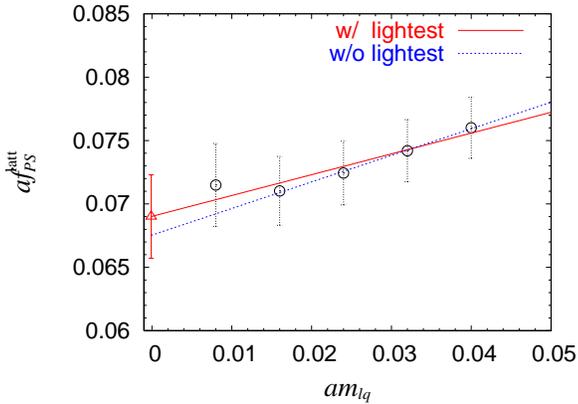}
  \vspace{-3ex}
  \caption{The chiral extrapolation of the pseudo-scalar decay constant
           at $am_{hq}$=0.4.}
  \label{fig:chiral-fD-040}
 \end{figure}
 The leptonic decay constants are obtained in a standard way from
 two-point functions with wall source and wall sink.
 Such correlation functions are still rather noisy, and the resulting
 decay constants depend on the fit ranges by as much as 10\%.
 In the following analysis, we thus adopt a set of fit ranges and
 take 10\% uncertainty as a systematic error.
 Figure \ref{fig:chiral-fD-040} shows the chiral behavior of the decay
 constant at $am_{hq}$=0.4.
 Since the data with $am_{lq}$=0.008 looks unnaturally high, we
 performed linear extrapolations with and without this data point.
 The difference between the two extrapolations are taken as a systematic
 error again.
 After interpolation to $am_{hq}$=$am_{\rm charm}$, we obtain
 \begin{eqnarray}
   && f_{D}^{\rm latt}     = 200(9)(^{+20}_{-21})\ {\rm MeV},\\
   && f_{D_{s}}^{\rm latt} = 216(7)(22)\ {\rm MeV},
 \end{eqnarray}
 where the systematic errors are summed up in quadrature.
 At the moment we quote the lattice bare values because we do not know
 how significant the $O(a^nm_{\rm charm}^n)$ error is.
 
 The SU(3) flavor breaking ratio of the decay constant is one of the
 quantities for which lattice QCD can attain a high precision because
 most of uncertainties are expected to cancel in the ratio.
 We can anticipate that the on-going charm experiments make it possible
 to measure this ratio at a few \% accuracy,
 and so this is a good quantity for which to compare the lattice result
 with the experimental one although the uncertainties in $m_s$, the
 chiral extrapolation and quenching effects must first be under
 control.
 Our current result is
 \begin{eqnarray}
   f_{D_{s}}/f_{D} = 1.08(3)(^{+2}_{-0}).
 \end{eqnarray}

\section{SUMMARY}

 We have started a exploratory study of massive DWF aiming at precision
 lattice calculations of the $D$ meson system.
 The preliminary results show no evidence of the DWF mechanism failing
 at large masses, while it may be better to improve how to impose a
 stopping condition.
 However, for precision calculations, we need to understand scaling
 violations, especially how serious the $O(a^nm_{\rm charm}^n)$ error
 is.
 We also need more statistics to have better control over chiral
 extrapolations.

\end{document}